# Utilizing Novel Quantum Counters for Grover's Algorithm to Solve the Dominating Set Problem

Jehn-Ruey Jiang
Department of Computer Science and Information Engineering
Naitonal Central University
Taoyuan City, Taiwan
jrjiang@csie.ncu.edu.tw

Qiao-Yi Lin
Department of Computer Science and Information Engineering
Naitonal Central University
Taoyuan City, Taiwan
111502563@cc.ncu.edu.tw

*Abstract*—Grover's algorithm is a well-known unstructured quantum search algorithm run on quantum computers. It constructs an oracle and calls the oracle $O(\sqrt{N})$ times to locate specific data out of *N* unsorted data. This represents a quadratic speedup compared to the classical unstructured data sequential search algorithm, which requires to call the oracle O(*N*) times. We are currently in the noisy intermediate-scale quantum (NISQ) era in which quantum computers have a limited number of qubits, short decoherence time, and low gate fidelity. It is thus desirable to design quantum components with three good properties: (i) a reduced number of qubits, (ii) shorter quantum depth, and (iii) fewer gates. This paper utilizes novel quantum counters with the above-mentioned three good properties to construct the oracle of Grover's algorithm to efficiently solve the dominating set problem (DSP), as defined below. For a given graph *G*=(*V*, *E*), a dominating set (DS) *D* is a subset of the vertex set *V*, such that every vertex is in *D* or has an adjacent vertex in *D*. The DSP is to decide for a given graph *G* and an integer *k* whether there exists a DS with size *k*. Algorithms solving the DSP have many applications. For example, they can be applied to check whether *k* routers suffice to connect all computers in a computer network. The DSP is an NP-complete problem, indicating that no classical algorithm exists to solve the DSP with polynomial time complexity in the worst case. Therefore, using quantum algorithms, such as Grover's algorithm, to exploit the potent computational capabilities of quantum computers to solve the DSP is highly promising. We execute the whole quantum circuit of Grover's algorithm using novel quantum counters through the IBM Quantum Lab service to validate that the circuit can solve the DSP efficiently and correctly.

*Keywords—Grover's algorithm, oracle, quantum circuit, quantum algorithm, dominating set problem*

I. INTRODUCTION

Quantum computers possess powerful computational capabilities through quantum mechanical phenomena, such as quantum superposition, entanglement, interference, and tunneling, and hence have gained widespread attention in recent years [1]. Unlike classical computers that perform computations based on bits, quantum computers perform computations based on quantum bits (or qubits). A bit is either 0 or 1; a qubit can simultaneously be in a superposition state representing both 0 and 1, and only collapses to a definite 0 or 1 state upon measurement. As a result, with the superposition of *n* qubits, we can represent and operate on all $2^n$ states at the same time. However, with *n* classical bits, we can represent and operate only one out of $2^n$ states of *n* bits at a time. The computation power of quantum computers thus scales exponentially with the number of qubits, surpassing the computational capabilities of classical computers that scale linearly with the number of qubits. This leads to quantum supremacy in the sense that quantum computers can achieve computation that classical computers can never achieve [2].

Numerous studies have proposed quantum algorithms run on quantum computers to effectively solve challenging problems. These quantum algorithms often exhibit significant acceleration compared to classical algorithms run on classical computers. For instance, the Deutsch-Jozsa algorithm [3], introduced in 1992, can determine whether a black-box function or an oracle of *N* input bits is a constant function or a balanced function. In contrast to classical algorithms that require $2^N$ oracle calls, the Deutsch-Jozsa algorithm only needs one oracle call, resulting in exponential speedup. Another example is Shor's algorithm [4], proposed in 1994, which can efficiently factorize semiprimes with polynomial time complexity. As a result, it can break RSA encryption systems whose security is based on semiprime factorization, providing exponential speedup compared to the fastest classical algorithm, the general number field sieve (GNFS) algorithm. Additionally, Grover's algorithm [5], introduced in 1996, relies on the concept of quantum amplitude amplification to solve unstructured data search problems with $O(\sqrt{N})$ oracle calls for *N* data items. This represents a quadratic speedup over classical sequential search algorithms for unstructured data.

This paper utilizes Grover's algorithm to solve a well-known NP-complete problem, the dominating set problem (DSP), as defined below. For a given graph *G*=(*V*, *E*), a dominating set (DS) *D* is a subset of the vertex set *V*, such that every vertex is in *D* or has an adjacent vertex in *D*. The DSP is to decide for a given graph *G* and an integer *k* whether there exists a DS with size *k*. Algorithms solving the DSP have many applications. For example, they can be applied to check whether *k* routers suffice to connect all computers in a computer network. Since the DSP is an NP-complete problem, no classical algorithm exists to solve the DSP with polynomial time complexity in the worst case. Therefore, using Grover's algorithm to exploit the potent computational capabilities of quantum computers to solve the DSP is highly promising.

We are currently in the noisy intermediate-scale quantum (NISQ) era [6] in which quantum computers have a limited number of qubits, short decoherence time, and low gate fidelity. It is thus desirable to design quantum components with three good properties: (i) a reduced number of qubits, (ii) shorter quantum depth, and (iii) fewer gates. This paper utilizes novel quantum counters with the three good properties to construct the oracle of Grover's algorithm to efficiently solve the DSP. For a given graph *G* and an integer *k*, Grover's algorithm can identify all DS's of *G* with size *k*, and thus solve the DSP. We execute the whole quantum circuit of Grover's algorithm using novel quantum counters through the IBM Quantum Lab service [7] to validate that the quantum circuit can solve the DSP efficiently and correctly.

The remaining sections of this paper are organized as follows. Section II introduces some preliminaries. In Section III, we describe the quantum circuit design of the novel quantum counter for the oracle of Grover's algorithm, whereas Section IV showcases the experimental results of executing the whole quantum circuit of Grover's algorithm through the IBM Quantum Lab service. Finally, Section V concludes this paper.

## II. PRELIMINARIES

### A. The Dominating Set Problem

Given an undirected graph $G(V, E)$ and a positive integer $k$, $0 < k \leq |V|$, the dominating set problem (DSP) is to determine whether there exists a subset $V'$ of $V$ such that $|V'|=k$ and for every vertex $v \in V$, $v \in V'$ or $u \in V'$, where $(u, v) \in E$. The subset $V'$ is referred to as a dominating set (DS) with size $k$. To sum up, the DSP is a decision problem, where the input consists of an undirected graph $G(V, E)$ and a positive integer $k$. The output of the problem is "true" if there exists a DS with size $k$ in the given graph $G$; otherwise, the output is "false". Fig. 1 illustrates all possible DS's of size 2 for an undirected graph with 6 vertices.

This paper, however, considers the extended version of the DSP instead of the original decision-version DSP. Specifically, for a given graph $G$ and a given integer $k$, the extended DSP considered in this paper is to identify all DS's of size $k$ in $G$.

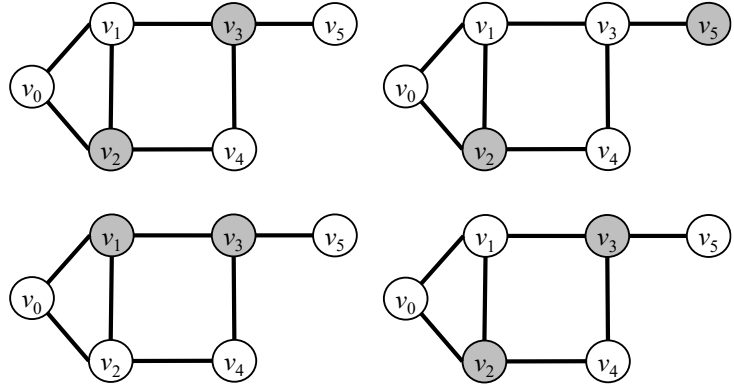

Fig. 1. All dominating sets with size 2 (of vertices in gray circles) of an undirected graph with 6 vertices.

### B. Grover's Algorithm

Grover's algorithm [5] is a quantum search algorithm designed to identify a specific data item, referred to as a target input instance, from unstructured data that are not arranged in a specific or structured order. Below we first introduce the concept of the oracle that is invoked by Grover's algorithm to identify the target input instance. An oracle or black-box function $f$: $\{0,1\}^n \rightarrow \{0,1\}$ takes $n$ bits as input, resulting in a total of $N=2^n$ possible state combinations. Among the combinations, there exists a target input instance $x^*$ to make $f(x^*)=1$, while every other input x makes $f(x)=0$, as expressed in the following equation.

$$f(x) = \begin{cases} 1, \text{if } x = x^* \\ 0, \text{if } x \neq x^* \end{cases} \quad (1)$$

Like other quantum algorithms, Grover's algorithm leverages the superposition of qubits. Initially, all $n$ input qubits are set to a uniform superposition, where each qubit is in a state of both $|0\rangle$ and $|1\rangle$ (may simply be denoted as 0 and 1 later) simultaneously. When measured, each qubit has a probability of 1/2 to be observed as state 0 and a probability of 1/2 to be observed as state 1. In total, the $n$ qubits exist in a superposition of all $N=2^n$ possible state combinations of $n$ qubits. This superposition allows each combination to be measured and read out with a probability of 1/$N$. Through the superposition of $n$ qubits, Grover's algorithm can perform operations on all possible input state combinations simultaneously. The key aspect of Grover's algorithm lies in the design of a phase oracle, denoted as $U_f$. If the input to the function is the target input instance $x^*$, the phase oracle performs a phase inversion on the input; otherwise, no operation is conducted, as expressed below.

$$U_f|x\rangle = \begin{cases} |x\rangle, \text{if } x = x^* \\ -|x\rangle, \text{if } x \neq x^* \end{cases} \quad (2)$$

Subsequently, Grover's algorithm performs a diffusion operation (or diffuser) on all input qubits through a diffuser. The purpose of the diffusion operation is to amplify the probability amplitudes of the target input instance $x^*$ to increase its probability density while decreasing the others'. As shown in Fig. 2 [8], the diffusion operation causes the probability amplitudes of all input instances to be inverted above the amplitude mean (denoted as $\mu$). Thus, positive amplitudes above $\mu$ are inverted downwards, becoming a little smaller, whereas the negative amplitude below $\mu$ that is caused by the phase oracle is inverted upwards, becoming significantly larger. The upper part of Fig. 2 illustrates the states of the qubits during the execution course of Grover's algorithm, the middle part represents the qubit states after the phase oracle has inverted the phase of the target input instance, and the lower part illustrates the qubit states after the diffusion operation.

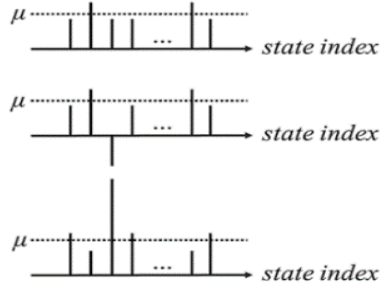

Fig. 2. The schematic diagram of the diffusion operation [8]

Grover's algorithm can also deal with the case of multiple target input instances. Fig. 3 illustrates the overall quantum circuit of Grover's algorithm. According to [9], Grover's algorithm needs to repeat the phase oracle and diffusion operation $\lfloor \pi/4\sqrt{N/M} \rfloor$ times to correctly identify the target input instance, where $N$ is the total number of possible state combinations of input qubits, and $M$ is the number of target input instances, $M \geq 1$. Typically, the quantum counting algorithm [10] can be employed to determine the number of target input instances, which will be described in the next subsection. Finally, Grover's algorithm performs quantum measurements. If $M=1$, the qubit state combination with the highest probability corresponds to the unique target input instance. If $M>1$, the qubit state combinations with significantly high probabilities correspond to target input instances.

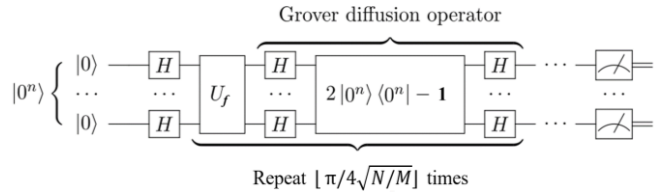

Fig. 3. The overall quantum circuit of Grover's algorithm (adapted from [8])

The literature presents numerous algorithms [11-20] based on Grover's algorithm to solve various NP-hard and NP-complete problems, including the maximum clique problem [11], k-coloring problem [12], chromatic number problem [13], list coloring problem [14], graphical game Nash equilibrium problem [15], Hamiltonian cycle problem [16-17], maximum satisfiability problem [18], exact cover problem [19], and vertex cover problem [20]. The readers are referred to [11-20] for details of these algorithms.

*C. The Quantum Counting Algorithm*

The quantum counting algorithm [10], proposed in 1998, aims to determine the number of target input instances or solutions to a given problem. It is based on the quantum phase estimation (QPE) algorithm and Grover's algorithm. Fig. 4, as presented in [21], provides an illustration of the quantum circuit for the quantum counting algorithm, with detailed descriptions outlined below.

In Fig. 4, Register 1 consists of $t$ qubits and serves as the counting qubits for the QPE algorithm. Generally, increasing the number $t$ of counting qubits allows for a more accurate estimation of the actual solution count $M$. Additionally, Register 2 is composed of $n+1$ qubits and serves as the input for the controlled Grover oracle (CGO) iterators which are represented as $G^{2^0}, G^{2^1}, \ldots, G^{2^{t-1}}$ in Fig. 4, corresponding to controlling the Grover oracle to repeat $2^0, 2^1, \ldots, 2^{t-1}$ times. The inverse quantum Fourier transform (denoted as $FT^\dagger$ in Fig. 4) is utilized to transform the counting qubits from the Fourier basis to the computational basis for later measurement. The measurement results of the counting qubits can be used to determine the phase (or phase rotation) corresponding to the eigenvalue of the Grover oracle. This phase information is in turn employed to deduce the number of target input instances of the Grover oracle.

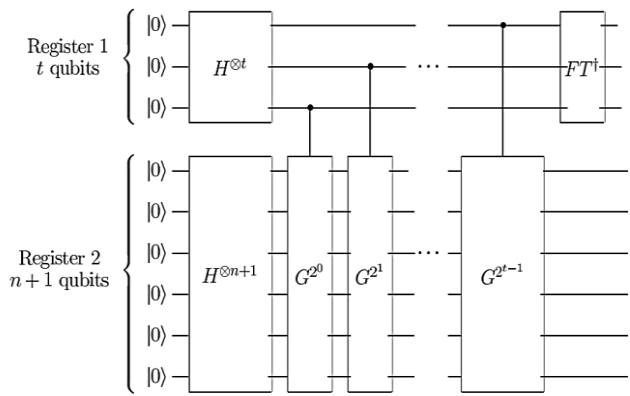

Fig. 4. Illustrative quantum circuit of the quantum counting algorithm [21]

III. THE PROPOSED QUANTUM CIRCUIT OF THE GROVER ORACLE TO SOLVE THE DOMINATING SET PROBLEM

This section describes the proposed quantum circuit of the Grover oracle to solve the DSP. If the size of the vertex set corresponding to the input state combination is exactly $k$, and this vertex set is a DS, then the oracle outputs $|1\rangle$; otherwise, the oracle outputs $|0\rangle$. The constraints of the DSP are further divided into two subconstraints, of which corresponding quantum circuit designs are described in the following subsections.

A. *Is the size of the vertex set equal to k?*

To check whether the size of a vertex set is $k$, we design a quantum counter circuit to determine the size of a vertex set, drawing inspiration from an existing quantum counter circuit proposed by Heidari et al. in [22]. The existing quantum counter is cyclic. Its quantum circuit is depicted in Fig. 5 and described below. The circuit features a control qubit denoted as $c$, along with $n$ counting qubits $a_0, a_1,…,a_{n-1}$, initialized as $|0\rangle^{\otimes n}$. The control qubit $c$ can take a series of control qubit states. For a state of $|1\rangle$ in the series, the quantum counter is increased by 1. On the contrary, for a state of $|0\rangle$ in the series, no operation is performed. However, if the quantum counter is already in the state $|1\rangle^{\otimes n}$, increasing it by 1 will make the state cycle back to $|0\rangle^{\otimes n}$. We observe that the existing quantum counter applies an X gate to a qubit after each controlled-not gate (CX gate), and applies $n-1$ X gates to total $n-1$ qubits after a multi-controlled not gate (MCX gate). This causes the problem of increasing the number of required quantum gates and expanding the quantum circuit depth.

To improve the existing quantum counter to mitigate its problem, this study proposes a novel quantum counter circuit design to be used in the Grover oracle, as shown in Fig. 6. The improvement involves rearranging the order of CX gates and the MCX gate to reduce the number of X gates applied. We have noted that the quantum circuit should undergo the transpiling process to be one with only fundamental gates to be executed on practical quantum computers. However, by observing the top-level quantum circuits depicted in Figs. 5 and 6 for comparing the improved quantum counter and the existing one, we infer that the improved quantum counter decreases the number of applied X gates by $2n-2$, and reduces the quantum circuit depth by $n$ for a quantum counter with $n$ counting qubits.

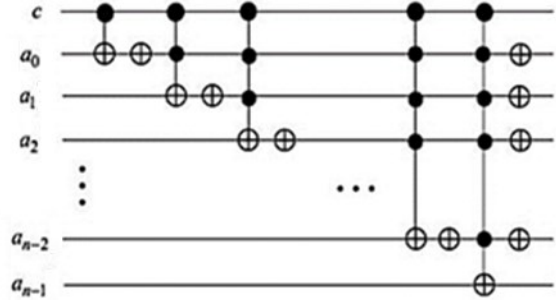

Fig. 5. The quantum circuit of the existing quantum counter proposed by Heidari et al. (adapted from [22])

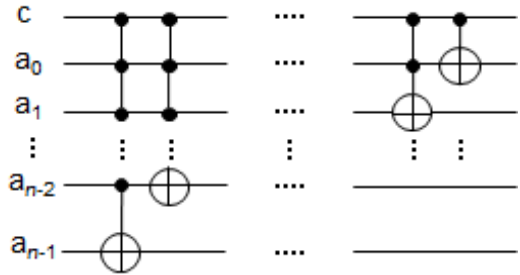

Fig. 6. The quantum circuit of the improved quantum counter used in this study

B. *Is every vertex dominated?*

A vertex $v$ is said to be dominated by a vertex set $V'$ if $v$ is in $V'$ or $v$ has an adjacent vertex in $V'$. This study utilizes a quantum checker that consists of a controlled quantum counter and an ancilla qubit to check whether every vertex is dominated by a vertex set. The quantum checker has a control qubit, several counting qubits, and an ancilla qubit. Every vertex has its corresponding checker. Thus, there are $n$ quantum checkers for a graph of $n$ vertices. Moreover, every vertex has its corresponding vertex qubit. A qubit corresponding to a vertex in a vertex set will be connected to the control qubit to activate the quantum checker corresponding to each of its neighbors. Hence, for a vertex that has $d$ neighbors, its corresponding qubit will be connected to $d$ quantum checkers, and there are $d$ vertex qubits connected to the control qubit of its checker. The quantum checker increases its quantum counter if the control qubit is $|1\rangle$. Thus, there should be $\lceil \log d \rceil$ counting qubits to keep the value of the quantum counter to range from 0 to $d-1$. The ancilla qubit is initialized to be $|1\rangle$. When the quantum counter value is $|00\rangle$, the ancilla qubit is flipped be to $|0\rangle$; otherwise, it is kept to be $|1\rangle$. In summary, if a vertex is dominated, the ancilla qubit of its associated quantum check has the state of $|1\rangle$; otherwise, the ancilla qubit has the state of $|1\rangle$. Fig. 7 shows the quantum circuit of the quantum checker with two counting qubits.

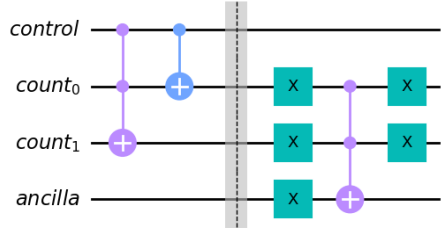

Fig. 7. The quantum circuit of the quantum checker with two counting qubits

## IV. EXPERIMENTAL RESULTS

This study adopts the quantum circuit architecture in Fig. 8 to realize Grover's algorithm. As shown in Fig. 8, H gates are first applied to input instance qubits to make them stay in state $|+\rangle$ to form the uniform superposition of states corresponding to all input instances. Moreover, some ancilla qubits and counting qubits are initialized to be in state $|0\rangle$, and one qubit is in the state $|-\rangle$. The unitary transformation $U_\omega$ is employed to mark qubits associated with the target input instances (later called target input instance qubits, for short) as $|1\rangle$. Subsequently, through the application of the MCX gate on the quantum state $|-\rangle$, phase kickback occurs to reverse the phase of the target input instance qubits. As will be shown later, the target input instance qubits may not be directly involved in the MCX gate, and $U_\omega$ includes many ancilla qubits and counting qubits initialized to $|0\rangle$. Hence, it is necessary to incorporate the conjugate transpose $U_\omega^\dagger$ of $U_\omega$ after the MCX gate to form a complete quantum oracle. The purpose of $U_\omega^\dagger$ is to perform uncomputation (i.e., the inverse transformation of $U_\omega$) to reset the ancilla and counting qubits used in $U_\omega$ back to the initial state $|0\rangle$, enabling the correct marking of target input instance qubits in the next iteration of Grover's algorithm and allowing the effect of phase kickback to be reflected on the target input instance qubits. Following the oracle, the diffuser is performed. Note that both the oracle and the diffuser must be repeated $\lfloor (\pi/4) \times \sqrt{N/M} \rfloor$ times before the final measurements to maximize the occurrence probability of the measurement results corresponding to the target input instances.

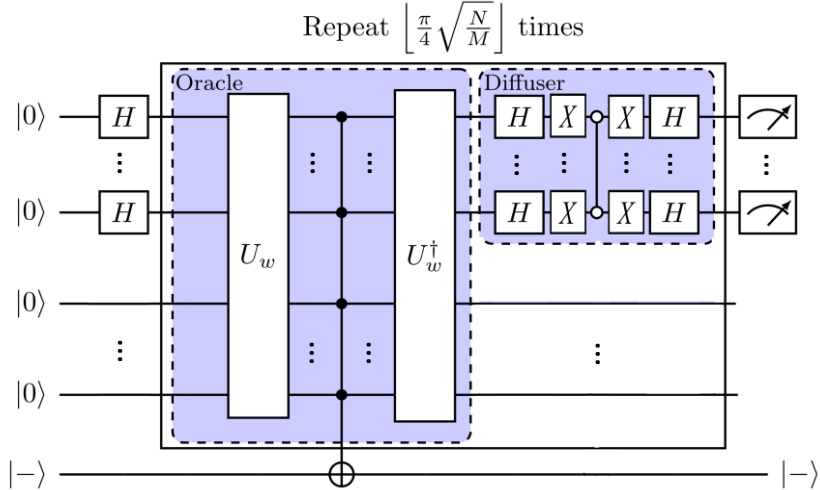

Fig. 8. The quantum circuit of Grover's algorithm using the phase kickback for reversing target input instance qubits

Below we illustrate the construction of the Grover oracle to solve the DSP with $k=2$ for the undirected graph $G=(V, E)$ with six vertices given in Fig. 1, where $V=\{v_0, v_1, v_2, v_3, v_4, v_5\}$ with $|V|=6$. Furthermore, we perform experiments through the IBM Quantum Lab service to simulate Grover's algorithm with the oracle using the improved quantum counters to validate the design correctness for solving the above-mentioned DSP.

Fig. 9 depicts the quantum circuit for a single iteration of Grover's algorithm applied to solving the DSP with $k=2$ for the undirected graph shown in Fig. 1. The quantum circuit of $U_\omega$, composed of the improved quantum counter, needs to be combined with its conjugate transpose quantum circuit of $U_\omega^\dagger$ to form a complete Grover oracle. As previously mentioned, the purpose of $U_\omega^\dagger$ is to reset the state of each ancilla qubit and counting qubit back to the initial state $|0\rangle$, ensuring the correct marking of the target input instance qubits in the subsequent iterations and enabling the effect of phase kickback to be reflected in the input instance qubits.

As shown in Fig. 9, the quantum circuit of Grover's algorithm includes 28 qubits, denoted as $q_0, \ldots, q_{27}$. The first six qubits $q_0, \ldots, q_5$ correspond to the input instance qubits or the vertex qubits associated with vertices $v_0, v_1, v_2, v_3, v_4,$ and $v_5$. They are with the initial state of $|+\rangle$, achieved by performing one H gate. They are input instance qubits in the uniform superposition of the states corresponding to all input instances. Every vertex has an associated quantum checker consisting of two counting qubits and one ancilla qubit. So total of 18 qubits, namely $q_6, \ldots, q_{23}$, are associated with quantum checkers. Moreover, three qubits $q_{24}, \ldots, q_{26}$ are counting qubits of a quantum counter to trace the number of vertices included in the vertex set. Note that the number of counting qubits is $\lceil \log |V| \rceil = \lceil \log 6 \rceil = 3$. Finally, the last qubit $q_{27}$ is the qubit with the initial state of $|-\rangle$, achieved by performing one X gate and one H gate. It is the target qubit of an MCX gate is used to realize phase kickback when the

MCX gate is activated (i.e., when all the control qubits are of the state $|1\rangle$). The control qubits of the MCX gate are qubits $q_8$, $q_{11}$, $q_{14}$, $q_{17}$, $q_{20}$, $q_{23}$, $q_{24}$, $q_{25}$, and $q_{26}$. Qubits $q_8$, $q_{11}$, $q_{14}$, $q_{17}$, $q_{20}$, and $q_{23}$ are respectively the ancilla qubit of a quantum checker. They are used to activate the MCX gate when vertex $v_0$, $v_1$, $v_2$, $v_3$, $v_4$, and $v_5$ are dominated, respectively. Also note that the inverse of counting qubit $q_{24}$ and counting qubit $q_{26}$, and counting qubit $q_{25}$ are control qubits of the MCX gate. If they are of the state $|010\rangle$, then the quantum counter reflects the value of $k=2$ and hence the MCX gate is activated. Note that the target qubit $q_{27}$ is of the state $|-\rangle$, thus phase kickback occurs when the MCX gate is activated.

It's essential to note that the combination of $U_\omega$ and $U_\omega^\dagger$ forms the complete Grover oracle. $U_\omega$ is realized by the quantum circuit between the first barrier and the barrier before the MCX gate with qubit $q_{27}$ as the target qubit. The construction of $U_\omega^\dagger$ involves reversing the order of all gates performed in $U_\omega$. In the undirected graph shown in Fig. 1, there are a total of 6 vertices, corresponding to 6 input qubits and resulting in a number $N = 2^6 = 64$ of possible state combinations. This study uses the quantum counting algorithm to determine the number $M$ of the solutions to the DSP for the graph in Fig. 1 is 4. Therefore, Grover's algorithm must repeat the complete Grover oracle and the diffuser for $\lfloor (\pi/4) \times \sqrt{N/M} \rfloor = \lfloor (\pi/4) \times \sqrt{64/4} \rfloor = 3$ times, followed by measurement, to make significant the occurrence probabilities of the measurement results corresponding to the target input instances. However, more significant occurrence probabilities are present if Grover's algorithm repeats the complete Grover oracle and the diffuser 4 times.

Fig. 10 displays the count histogram of measurement results of Grover's algorithm to solve the DSP for graph $G$ shown in Fig. 1 with $k = 2$. The algorithm is executed 1000 times (or shots) to produce the histogram with 4 times of oracle and diffuser repetitions. Note that the count histogram rather than the probability histogram is adopted, as the former can show experimental results more clearly than the latter. The measurement result counts for '001001', '001010', '001100', and '100100' are 189, 190, 206, and 308, respectively, which are significantly higher than the counts for other input state combinations. By converting the measurement results with significantly high counts into solutions, we have that '001001' represents a vertex set of $v_0$ and $v_3$, '001010' represents a vertex set of $v_1$ and $v_3$, '001100' represents a vertex set of $v_2$ and $v_3$, and '100100' represents a vertex set of $v_2$ and $v_5$, which indeed are all dominating sets of $G$ with size 2. In summary, graph $G$ has four dominating sets of size 2, and Grover's algorithm using the proposed quantum circuit designs can correctly identify them.

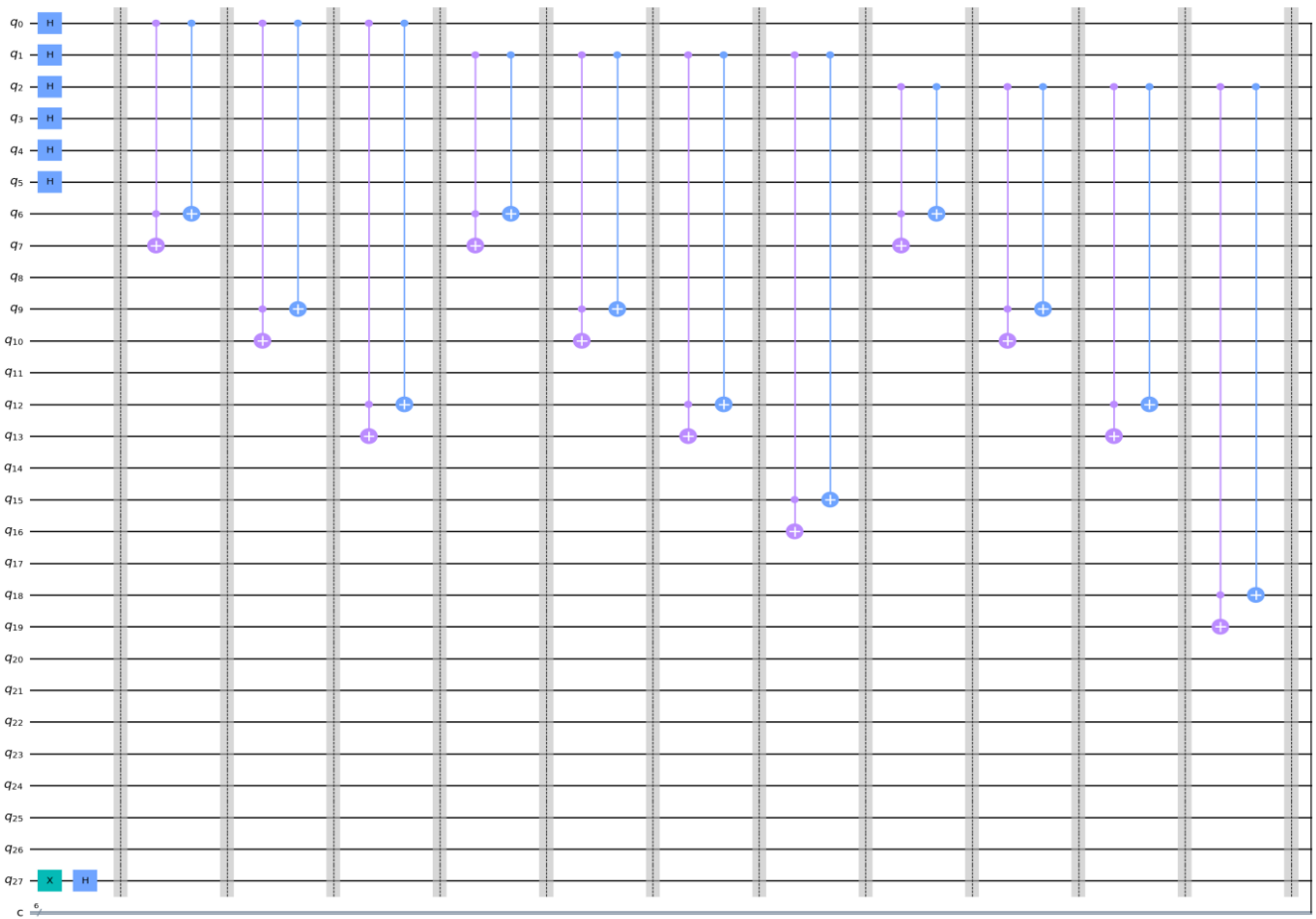

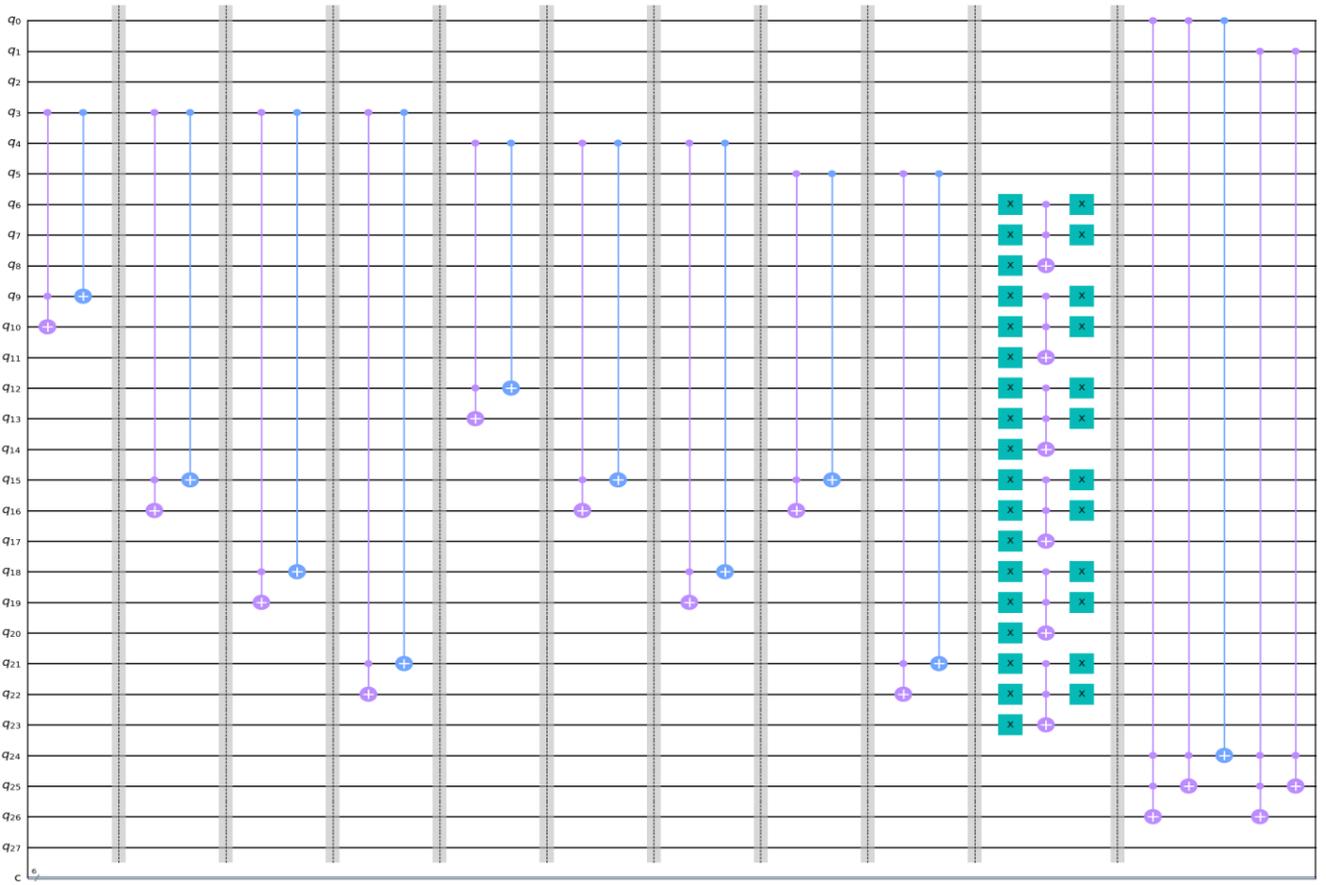
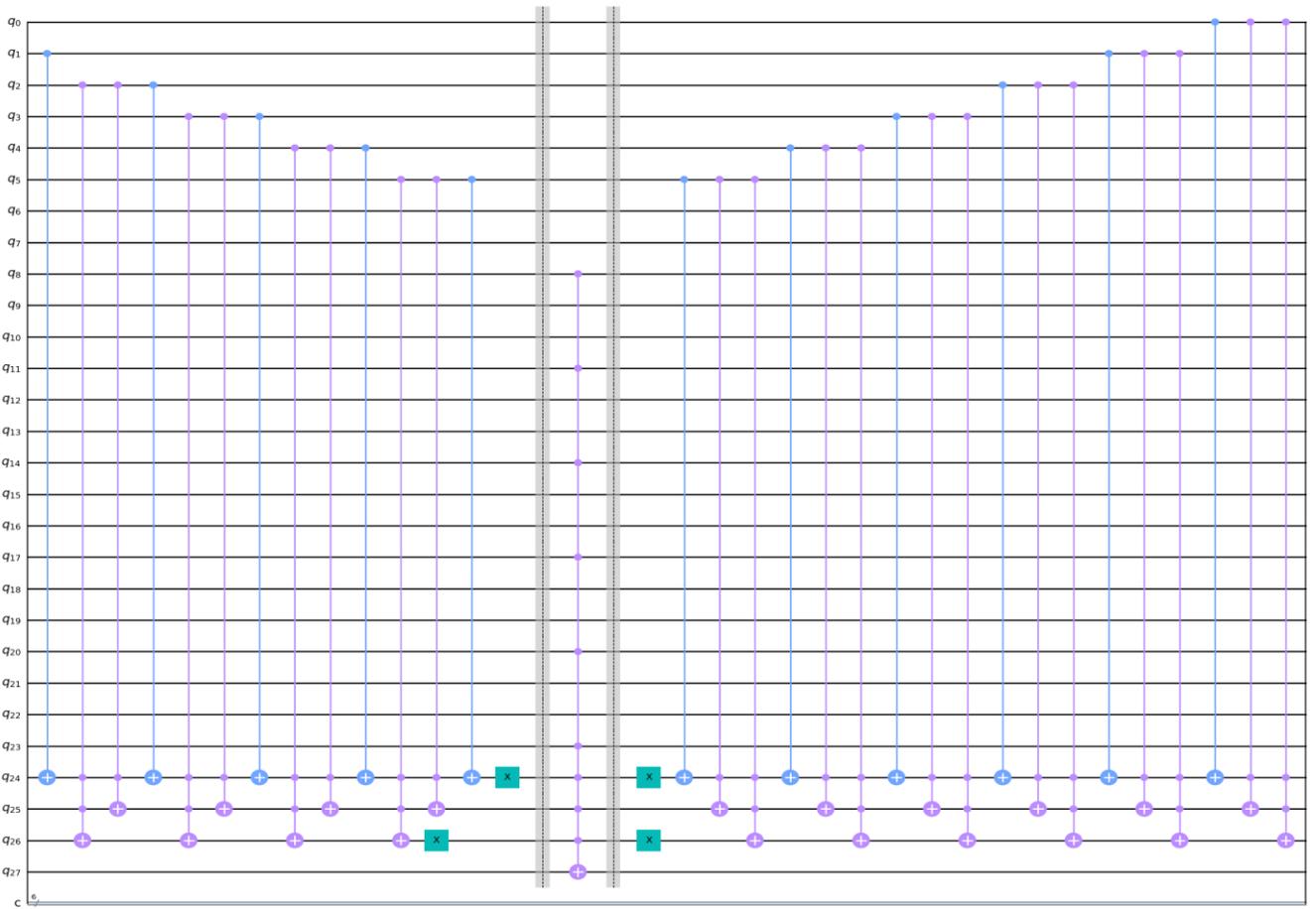

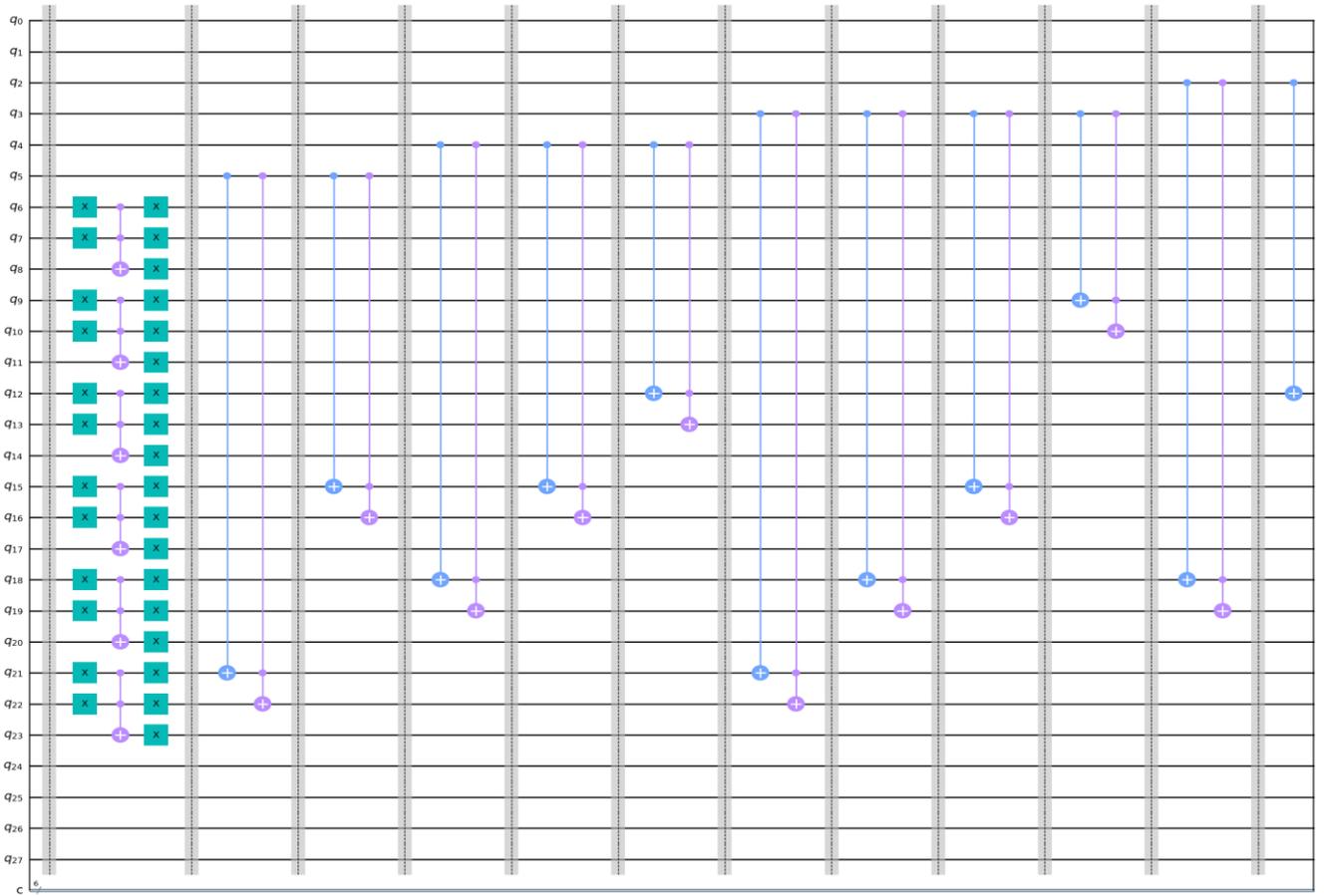
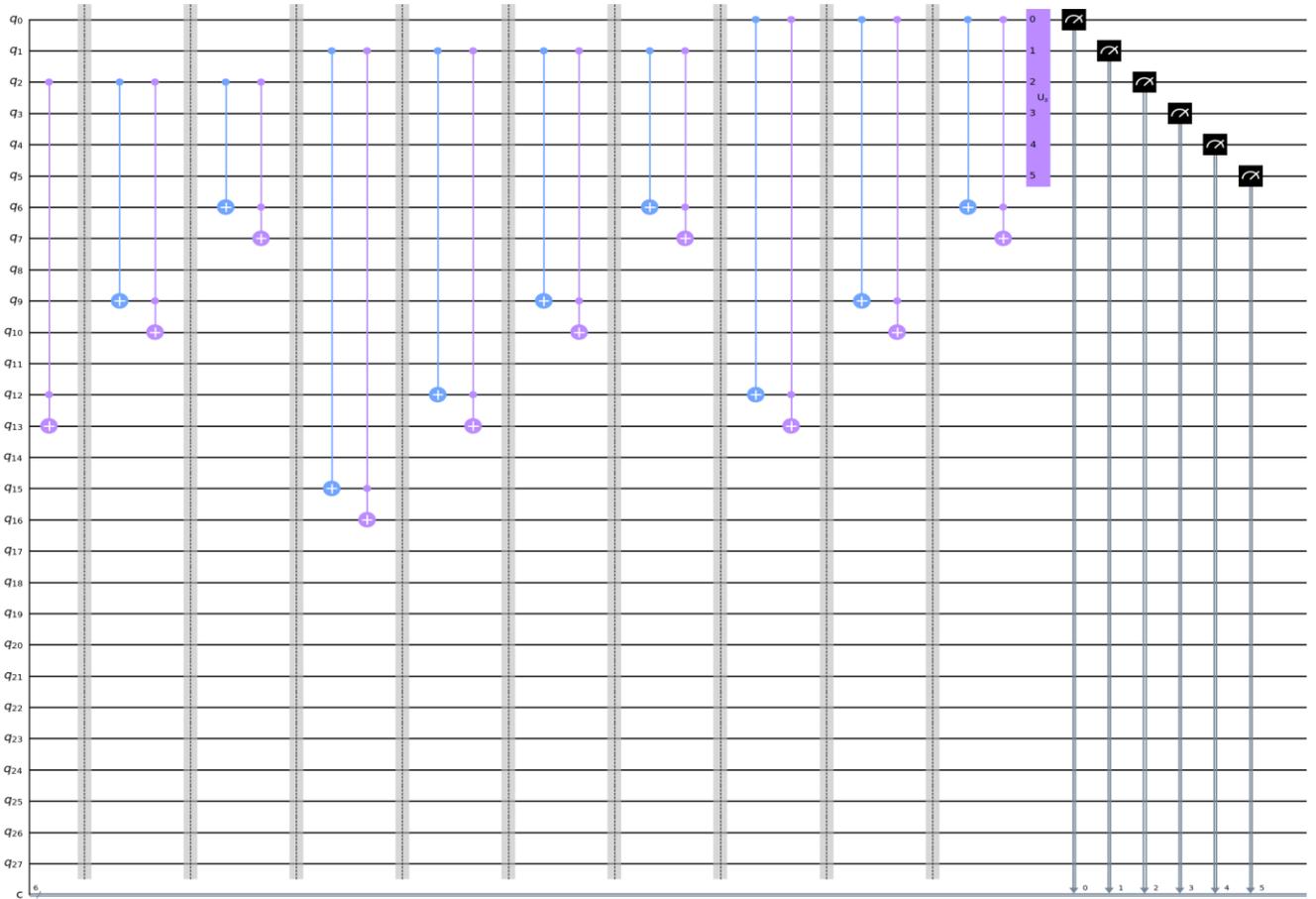

Fig. 9. The quantum circuit with a single repetition of the oracle and the diffuser of Grover's algorithm to solve the DSP with $k=2$.

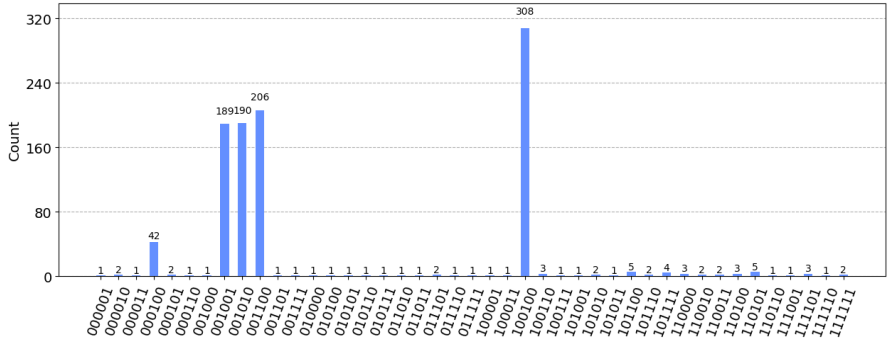

Fig. 10. The count histogram of the measurement results of Grover's algorithm to solve the DSP with $k = 2$.

## V. Conclusion

This paper utilizes a novel quantum counter to improve the oracle of Grover's algorithm to solve the DSP for a given graph and a given integer. Grover's algorithm can solve the DSP for a graph with $n$ vertices by calling the oracle $O(\sqrt{2^n})$ times, which represents a quadratic speedup compared to the classical unstructured data sequential search algorithm, which requires $O(2^n)$ oracle calls. The novel quantum counter makes the oracle have (i) a reduced number of qubits, (ii) a shorter quantum depth, and (iii) fewer gates. The whole quantum circuit of Grover's algorithm is simulated via IBM Quantum Lab services to validate its correctness. Note that when we simulate the quantum counting algorithm, we encounter some problems and the algorithm cannot run properly due to the large size of its quantum circuit. We assume that the quantum counting algorithm can return the exact number of solutions. Fortunately, these problems do not influence the correctness of Grover's algorithm. In the future, we plan to solve the problems. We also plan to compare the performance of Grover's algorithm with those of related classical algorithms for solving the DSP, such as the iterated greedy algorithm [23].

## Acknowledgment

We express our gratitude to the "National Taiwan University - IBM Quantum Computing Center (IBM Q Hub at NTU)" for providing us with access to the IBM Q system.